\documentclass{IEEEmce}

\usepackage[colorlinks,urlcolor=blue,linkcolor=blue,citecolor=blue]{hyperref}
\usepackage{upmath}
\usepackage[linesnumbered,ruled,vlined]{algorithm2e}
\usepackage{graphicx}
\usepackage{multirow}
\usepackage{textcomp}
\usepackage{scalerel}
\usepackage{graphics}
\usepackage[utf8]{inputenc}
\usepackage{amsfonts}
\usepackage[fleqn]{amsmath}
\usepackage{tikz}
\usepackage{tablefootnote}
\usepackage{threeparttable}
\usepackage{array}
\usepackage{multirow}
\usepackage{booktabs}
\usepackage{color}
\usepackage{array}
\usepackage{hhline}
\usepackage{soul}
\usepackage{comment}
\usepackage{tablefootnote}
\usepackage{url}
\usepackage{subcaption}
\usepackage{verbatim}

\jvol{XX}
\jnum{XX}
\paper{8}
\jmonth{xxx/xxx}
\publisheddate{00 xxxx 0000}
\currentdate{00 xxxx 0000}
\jname{IEEE Design \& Test}
\pubyear{2023}
\doiinfo{DNT.2023.Doi Number}

\begin{document}
\editor{Editor: Name, xxxx@email}
\bstctlcite{IEEEexample:BSTcontrol}
\title{Impact of Orientation on the Bias of SRAM-Based PUFs}

\author{Zain~Ul~Abideen}
\affil{Department of Computer Systems, Centre for Hardware Security, Tallinn University of Technology, 12616, Tallinn, Estonia}

\author{Rui Wang}
\affil{Intrinsic ID, HTC 83, 5656AG, Eindhoven, Netherlands}

\author{Tiago Diadami Perez}
\affil{Department of Computer Systems, Centre for Hardware Security, Tallinn University of Technology, 12616, Tallinn, Estonia}

\author{Geert-Jan Schrijen}
\affil{Intrinsic ID, HTC 83, 5656AG, Eindhoven, Netherlands}

\author{Samuel Pagliarini}
\affil{Department of Computer Systems, Centre for Hardware Security, Tallinn University of Technology, 12616, Tallinn, Estonia}
\markboth{}{Paper title}


\maketitle

\enlargethispage{10pt}

\chapterinitial{Physically Unclonable} Function (PUF) modules are a useful hardware security primitive due to their uniqueness, non-reproducible and unclonable features. Generally speaking, a PUF is used for two applications: secret key generation for cryptographic use and/or device authentication/provenance of Integrated Circuits (ICs) \cite{PUF_intro2}. PUFs exploit process variation (e.g., gate oxide thickness, size, and threshold voltage) that occurs naturally during the fabrication of ICs. Despite the fact that ICs are fabricated from identical layouts, every transistor presents slightly random electric properties which aid to generate a unique identity \cite{PUF_intro1}. Importantly, these random properties cannot be replicated even if an adversary has access to the full design, thus it provides a unique advantage for anti-counterfeiting measures \cite{PUF_authentication}. 

SRAM-based PUF technology is, by far, the most studied form of a PUF \cite{PUF_SRAM1, PUF_SRAM3, PUF_L2, PUF2021, PUF_2020, PUF_2012, PUF_2020_new}. The digital signature of an SRAM-based PUF is the raw entropy of the SRAM array, which is converted into a string of bits. SRAM-based PUFs offer a combination of simplicity, low cost, high reliability, scalability, and cryptographic strength, making them a preferred choice for commercial PUF solutions.  
Another advantage of SRAM-based PUFs is that they rely on standard SRAM IP that is typically available to designers. The same memory macros utilized for design purposes can also serve as a PUF, there is no need to have a customized memory macro. For these reasons, our work also focuses on SRAM-based PUFs.

In this work, we explore the impact that \emph{designer's choice} have on the bias pattern of SRAM-based PUFs.
We designed and fabricated a 65nm CMOS chip featuring eleven SRAMs with varying numbers of addresses, number of words, aspect ratio, and chosen bitcell. More importantly, we also considered the orientation of the memories in our evaluation. This study makes a significant contribution by exploring and analyzing the phenomenon of the bias pattern in SRAM-based PUFs. We aim to deepen our understanding of how orientation influences the observed bias patterns in SRAM-based PUFs.

We have updated the last paragraph of the introduction to highlight the contribution for the advancement in the field of SRAM-based PUF.


\section{Background} \label{sec:PUF_structure}
The main component of an SRAM memory is the bitcell, which consists of two CMOS inverters interconnected in a positive feedback loop, forming a bistable storage element. The initial state of each bitcell is determined by the process variation that arises during the IC's manufacturing. The stability of each bit is contingent upon the degree of threshold voltage mismatch between the local devices. The typical 6T-SRAM cell exhibits a preferred state due to stochastic variations in the threshold voltages of its transistors. This randomness in the initial values of 6T-SRAM results in a largely unpredictable, largely repeatable, and yet exclusive pattern of zeros and ones.

\begin{figure*}[th]
\centering \footnotesize
\includegraphics[width=1.0\linewidth]{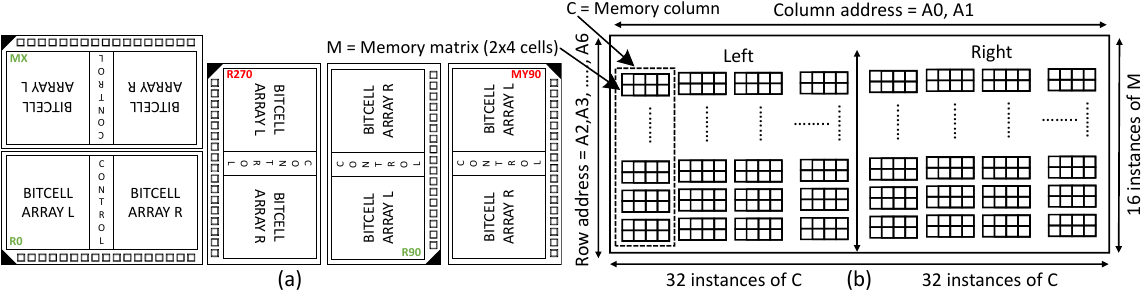}
\caption{Architecture and orientations of memory: low speed memory with 32-bit datawidth and 1024 location depth.}
\centering
\label{fig:memory_orientation}
\end{figure*}

The architecture of the memory and different valid orientations are illustrated in Fig. \ref{fig:memory_orientation}. All SRAM memories are single-port and utilize 6 transistors per bitcell. Furthermore, we utilized two distinct types of memories. Those with a bitcell size of approximately $\sim$0.65$\mu m^2$ are labelled as low density and high speed, whereas those with a bitcell size of approximately $\sim$0.50$\mu m^2$ exhibit high density but low speed. The high-speed memory adopts Standard-vt for both periphery and bitcells, while the low-speed memory utilizes Mixed-vt for periphery and High-vt for bitcells. 

SRAM macros can be internally arranged in a variety of ways. These are decisions that the memory compiler automatically performs given a user specification (i.e., desired number of addresses and datawidth). As shown in Fig. \ref{fig:memory_orientation} (a), the memories generated by a commercial SRAM compiler have half of the bits on the right and the other half on the left. The control circuitry lies in the centre. This arrangement is identical for high-speed and high-density variants. 

These large bitcell arrays are built by replicating a Memory matrix of size $m \times n$ repeatedly to form a larger matrix of size $M \times C$, as depicted in panel (b) of Fig. \ref{fig:memory_orientation}. By deciding values for $m$, $n$, and M, the memory compiler determines the aspect ratio of the memory and how many bitcells have to be muxed together. Let us consider an example with a datawidth of 64 bits and a depth of 128 locations, i.e., a memory that contains 8Kbits. The address A is 7 bits long, from which \{A0, A1\} index the columns while \{A2, A3, A4, A5, A6\} index the rows. 



The column mux ratio $m$ plays a crucial role in a memory array as it represents the number of memory cells that are connected to a shared bitline. The choice of $m$ involves a trade-off between density, access time, and power consumption. A higher column mux ratio increases memory density because more cells can be packed into a given area. However, it also introduces larger capacitance and longer bitline lengths, leading to slower access times and potentially higher power consumption. On the other hand, a lower column mux ratio improves access time.

The location of the SRAM macro within the chip is determined by the overall floorplan. In our test chip, we have utilized 11 SRAMs with different possible orientations, as depicted in Figure \ref{fig:memory_orientation}. The default orientation is denoted as R0, representing a rotation angle of zero degrees. The notation MX signifies mirroring along the x-axis, R270 refers to a rotation of 270 degrees, R90 represents a rotation of 90 degrees, and MY90 signifies mirroring along the y-axis followed by a rotation of 90 degrees (all rotations mentioned here are anti-clockwise).

In SRAM-based PUFs, the combination of bitcells forms the response data. This response data can be evaluated by a number of metrics. The \emph{reliability} metric measures the ability of a PUF to consistently reproduce its output response, independent of temperature variations and fluctuations in operating voltage. An ideal PUF should produce the same output in response to a given challenge, regardless of the environmental and operational conditions. The reliability of a PUF can be assessed by evaluating its within class hamming distance (WCHD), which is the fractional hamming distance between measurements taken at various conditions during reconstruction and a reference initial measurement (enrollment). 

SRAM-based PUFs leverage \emph{entropy} from process variations to build unique fingerprints for each identically fabricated chip. For \emph{entropy}, one important parameter is the \emph{bias pattern} that is linked with the PUF's physical characteristics. The response data often displays a consistent bias towards either a logical zero or one at given intervals. This bias cannot be easily determined by calculating the fractional hamming weight, which measures the percentage of ones in the raw output response. This recurrent phenomenon is usually referred to as the \emph{bias pattern}.

The biased SRAM-based PUFs will show higher auto-correlation values compared to non-biased devices. We select the SRAM-based PUF with the highest auto-correlation per SRAM-based PUF's design and run Fast Fourier Transform (FFT) on the auto-correlation. The auto-correlation has some periodic effect, and the period corresponds to the width of the bias pattern. 
Entropy in SRAM-based PUF is derived by inserting the Masked Hamming Weight (MHW) into min-entropy formula, and MHW is calculated through the percentage of ones in the output response after an XOR operation with the bias pattern \cite{min_entropy}. 

\subsection{ASIC Demonstration} \label{subsec:asic_demo}
\begin{figure*}
     \centering
     \begin{subfigure}[b]{0.49\textwidth}
         \centering
         \includegraphics[width=\textwidth]{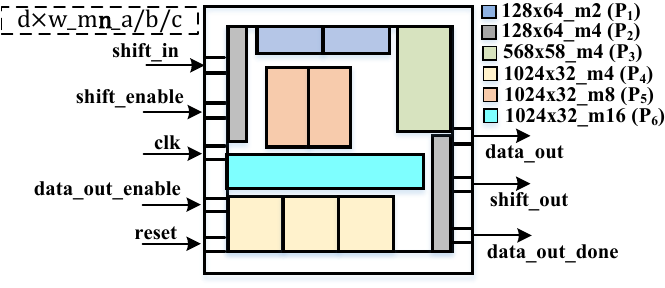}
         \caption{The simplified architecture of the chip.}
         \label{fig:PUF_floorplan_arc}
     \end{subfigure}
     \begin{subfigure}[b]{0.49\textwidth}
         \centering
         \includegraphics[width=\textwidth]{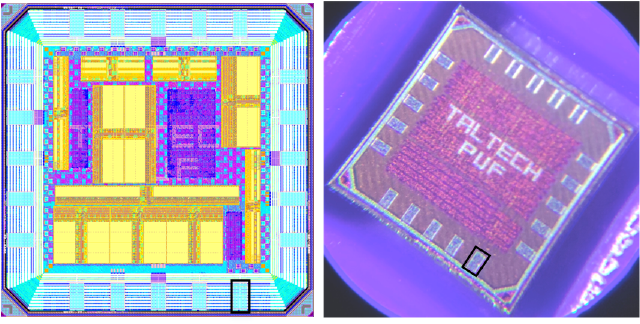}
         \caption{Layout (left) and die micrograph (right) of the fabricated chip. The highlighted pin marks the lower-right corner.}
         \label{fig:PUF_floorplan}
     \end{subfigure}
        \caption{The simplified architecture and chip design of the SRAM-based PUFs.}
\end{figure*}

The simplified internal architecture of the chip is depicted in Fig. \ref{fig:PUF_floorplan_arc}. The chip contains 6 distinct SRAM-based PUFs, some of which are replicas denoted by underscored letters (a, b, c). This results in a total of 11 SRAM-based PUFs within the chip \footnote{d denotes the depth or number of addresses, w denotes the data width, and m denotes the column mux ratio.}. The designed chip consists of a simple serial interface and eleven different SRAM-based PUFs.  
At the input of the serial interface, 15 bits are employed for the fully addressing and selecting any address of the eleven SRAM-based PUFs. 
At the output of the serial interface, 70 bits are retrieved, including 64 bits of data along with 3 start bits and 3 stop bits for data alignment. 
The chip includes both slow and fast memories, with the smaller memories ($128 \times 64$) being classified as fast memories while the rest of the memories are classified as slow. We have manually placed all memories multiple times to maximize the area usage. Next, we executed P\&R to generate the final layout of the chip. The control logic inside the chip is minimal while the majority of the area is occupied by memories, making buffers, inverters, and sequential cells less than 5\% of the circuit. The number of buffers, combinational cells, inverters, and sequential cells are 233, 640, 881, and 54, respectively.



Fig. \ref{fig:PUF_floorplan} shows the bare die in the right panel and its layout is illustrated in the left panel. One can recognize the placement of memories and IO cells as illustrated in the layout. A black rectangle has been added to facilitate the identification of the lower right corner of the chip. To validate the design, we have packaged 50 samples of the chip and all of the packaged samples were confirmed to have full functionality. We have designed a custom printed circuit board (PCB) for the purpose of testing our ASIC prototype which contains a chip socket, relays, and passive components to assist with measurements and filtering noise from the power supplies. The chip is controlled by a Raspberry Pi 3 Model B to facilitate the testing. The relays are used to turn on and off the VDD (1.2V) and VDDIO (3.3V) power supplies.


For the experiments, we recorded the PUFs' responses from the chips by power cycling every chip 10 times. The purpose of using relays became clear as we utilized them to turn on and off the chip to complete one power cycle. 
The Raspberry Pi sends serial bits to the chip and stores the corresponding PUF response in a text file. The Pi repeated this process by turning on and off the chip for the subsequent experiments. We repeat this experiment ten times to analyze the stability of the PUFs' response. In total, we collected 262,320 per reading per chip. Since we have 50 chips and perform power cycling 10 times, the total number of datapoints collected is 131M.


Thanks for highlighting this issue. We have updated figures 3 and 4. Now the axis and the fonts are visible.

\section{Results and Discussions} \label{sec:metrics_results}
Prior to presenting our bias-related analysis, we will examine three important parameters: WCHD, MHW, and entropy. These parameters provide valuable information on the behavior and characteristics of the SRAM-based PUF. To start this, we merged data from 50 chips and evaluated the corresponding metric for each SRAM measurement. A summary of the results is presented in Table \ref{tab:results} where the first column lists the type of memory, and the subsequent columns are WCHD, MHW, and entropy. The WCHD values, which are below 10\%, indicate a high level of reliability in the SRAM-based PUFs. Ideally, the MHW value should be around 0.5, and the results in the second column of Table \ref{tab:results} closely align with the ideal value. The entropy values are derived from the MHW values and also align well with the expected entropy levels. Furthermore, the entropy values obtained in our study were found to be comparable to those reported in \cite{PUF_2012}. The entropy values reported in \cite{PUF_2020_new} are slightly higher but their analysis is based on an older technology node, making a fair comparison challenging. 
Overall, the results confirmed that our SRAM-PUFs behave as expected and in line with previous studies.

\begingroup
\setlength{\tabcolsep}{2.0pt} 
\renewcommand{\arraystretch}{1.1} 
\begin{table*} [!htb]
\footnotesize \centering
\caption{Results for the robustness evaluation of SRAM-PUFs}
\label{tab:results}
\begin{tabular}{|p{2.2cm}|p{1.5cm}|p{1.4cm}|p{2.2cm}|p{2.5cm}|p{1.7cm}|p{0.8cm}|} \hline
\textbf{SRAM-PUF} & \textbf{WCHD (\%)} & \textbf{MHW} & \textbf{Entropy by one-probability} & \textbf{Bias pattern} & \textbf{Orientation} & \textbf{BD} 
\\ \hline
P\textsubscript{1}\_{a}, P\textsubscript{1}\_{b} & 5.8-8.5 & 0.391-0.622 & 0.685-1 & \parbox{1.5cm}{$0(32)1(64)0(64), ...$} & R0, R0 & +, + \\ \hline 
P\textsubscript{2}\_{a}, P\textsubscript{2}\_{b} & 5.8-8.8 & 0.440-0.564 & 0.826-1 & $0(32)1(64)0(64), ...$ & R90, R270 & +, - \\ \hline
P\textsubscript{3} & 5.5-7.0 & 0.430-0.539 & 0.811-1 & $0(29)1(29), ...$ & R270 & - \\ \hline
P\textsubscript{4}\_{a}, P\textsubscript{4}\_{b}, P\textsubscript{4}\_{c} & 5.0-9.1 & 0.435-0.541 & 0.824-1 & $0(16)1(16), ...$ & MX, MX, MX & +, +, + \\ \hline
P\textsubscript{5}\_{a}, P\textsubscript{5}\_{b} & 5.2-8.0 & 0.430-0.575 & 0.798-1 & $0(16)1(16), ...$ & R270, MY90 & -, - \\ \hline
P\textsubscript{6} & 5.1-7.0 & 0.390-0.580 & 0.713-1 & $0(16)1(32)0(32), ...$ & R0 & + \\ \hline
\end{tabular}
      \centering
\end{table*}
\endgroup

\begin{figure}[tb]
\centering \footnotesize
\includegraphics[width=1.0\linewidth]{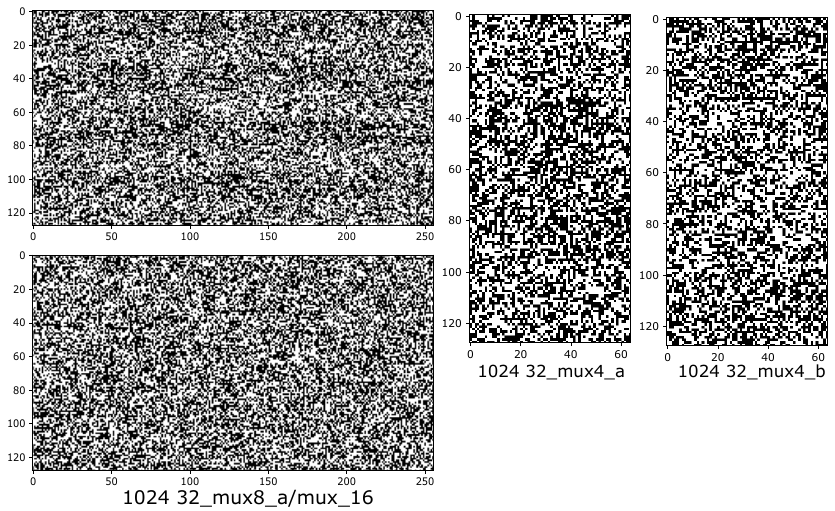}
\caption{The start-up pattern of the P\textsubscript{5}\_a, P\textsubscript{6}, P\textsubscript{4}\_{a} and P\textsubscript{4}\_{b} SRAM-based PUFs is represented by a sequence of bits, with white spaces denoting a logical one.}
\centering
\label{fig:PATTERN_SRAM_1024_32_mux16}
\end{figure}

\begin{figure}
     \centering
     \begin{subfigure}[b]{0.235\textwidth}
         \centering
         \includegraphics[width=\textwidth]{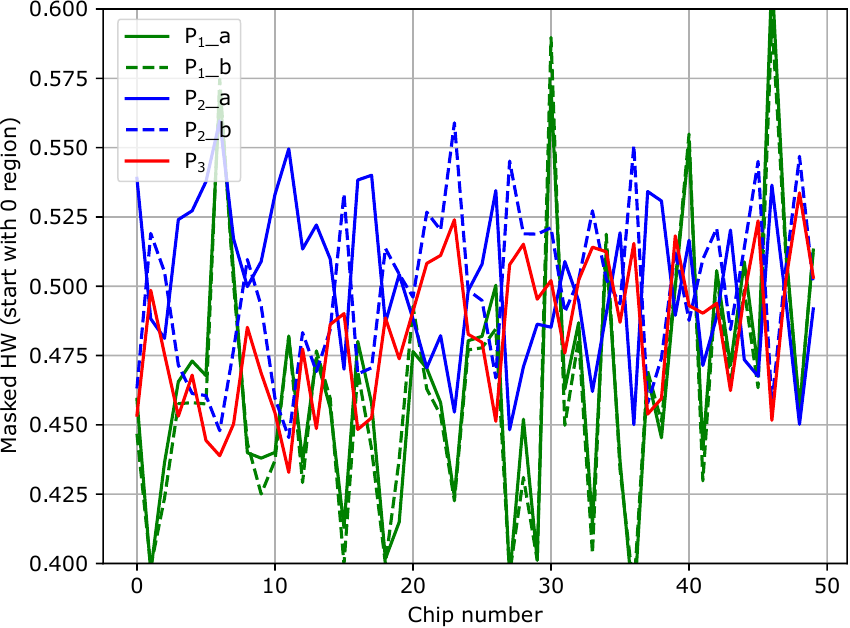}
         \label{fig:biasing(a)}
     \end{subfigure}
     \begin{subfigure}[b]{0.235\textwidth}
         \centering
         \includegraphics[width=\textwidth]{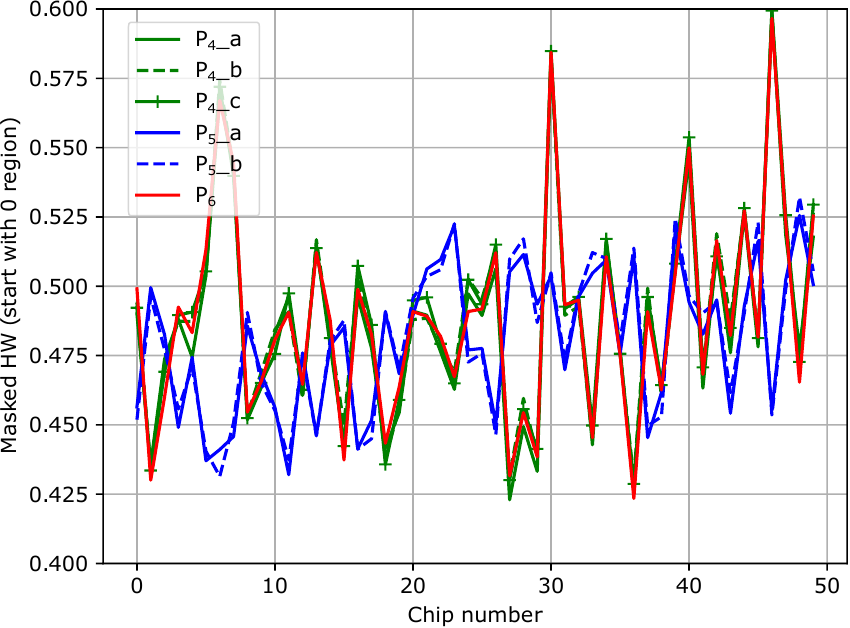}
         \label{fig:biasing(b)}
     \end{subfigure}
        \caption{The correlation of SRAM-based PUFs for the biasing pattern with baseline SRAM-based PUF (P\textsubscript{1}\_{a}).}
        \label{fig:biasing}
\end{figure}

Next, the analysis seeks to determine the impact of bias patterns on the SRAM-based PUFs over varying sizes, mux selection ratios, memory types and orientations. We found that, for this chip, the location of the memory on the floorplan has no impact on its functionality. However, when deciding how to arrange the floorplan, a designer may choose to rotate a memory to obtain better routability. Indeed, memory orientation affects the bias direction (BD). To start the analysis, the start-up pattern of SRAMs P\textsubscript{5}\_a, P\textsubscript{6}, P\textsubscript{2}\_a and P\textsubscript{2}\_b are plotted in Fig. \ref{fig:PATTERN_SRAM_1024_32_mux16}. One can visualize the biasing patterns that we will explain in the following results. The response vectors are concatenated into valid binary data from each row into a one-dimensional vector to determine the bias pattern. Next, we compute each auto-correlation for each chip (but only on the first measurement). The correlation between the MHW for all 50 chips is depicted in Fig. \ref{fig:biasing}. Notably, the direction of correlation varies from one SRAM-based PUF to another when considering P\textsubscript{1}\_a as the baseline. For example, if we compare the correlation peaks for the baseline and P\textsubscript{2}\_b in Fig. \ref{fig:biasing} then the peaks are opposite; thus, we report the bias direction as negative. A summary of the relationship between the memory orientation and the bias correlation direction is listed in the last two columns of the Table. \ref{tab:results}. 

\emph{Observation 1:} The bias pattern between different sizes is influenced by the data width, but it does not affect the direction of the bias pattern. The data width (word size), bias pattern, and the number of instances for different SRAM-based PUF instances are listed in the last three columns of Table \ref{tab:results}. For instance, consider two identical memories, P\textsubscript{2}\_{a} and P\textsubscript{2}\_{b}. Despite their identical nature, these memories exhibit different bias directions. One memory may have a positive bias direction, while the other may have a negative bias direction.

\emph{Observation 2:} Overall, the change in mux ratio causes the different aspect ratios, which have different bias patterns from 1024$\times$32\_mux8 and 1024$\times$32\_mux16. So, the larger mux ratio has an impact on the bias pattern. The identical SRAM-based PUFs have an identical bias pattern. On the other hand, the column mux ratio does not influence the direction of the bias pattern.

\emph{Observation 3:} The use of fast and slow memories do not have any connection with the width of the bias pattern. Additionally, they are also not related to the direction of the bias pattern. In other words, the use of different bitcells does not translate into different bias widths or directions.

\emph{Observation 4:} However, it is important to note that SRAM-based PUFs may display an alternating bias pattern, which is a result of the internal structure of SRAMs. As discussed in Section \ref{sec:PUF_structure}, the SRAM macro consists of two halves: one on the left and the other on the right. This arrangement leads to an interesting observation: the initial 32 bits of P\textsubscript{1}\_{a} and P\textsubscript{1}\_{b} tend to skew towards zero, followed by an alternating pattern of 64 bits.


\emph{Observation 5:} It is confirmed that two memory orientations, namely R270 and MY90, show a negative biasing direction when compared to the other memories. 

\emph{Observation 6:} It has been confirmed that the bias pattern direction observed in SRAM-based PUFs is not influenced by factors such as power planning and the overall floorplan of the chip, except for considerations specifically related to orientation. 

\emph{Observation 7:} Intra-die process variation exhibits uniqueness between individual dies, and thus, it does not contribute to the direction of the bias pattern. We further confirm that all chips come from the same wafer and have not been rotated on the MPW reticle.

According to observations 1-7, the direction of the bias pattern cannot be affected by varying sizes, mux ratios, memory types, memory structure, or process variations. The orientations are the only factors that determine it. We make the following hypothesis to potentially explain the direction of the bias pattern.
 
\emph{Hypothesis 1:} This effect emanates from the orientation of the bitcells themselves. The R90 orientation of the SRAM positions the bitcells vertically compared to the R0 orientation. When comparing the R90 orientation as a reference, the R270 and MY90 orientations flip the left and right bitcell arrays as illustrated in Fig. \ref{fig:memory_orientation}. This means that the placement direction of the bitcells associated with the first address of the SRAM becomes reversed. In other words, the direction of the bitcell placement linked to the first address of the SRAM becomes the opposite in these orientations. 


\emph{Hypothesis 2:} Understanding the impact of doping variations on the bias pattern is crucial for analyzing and mitigating their effects in SRAM-based PUFs. During the lithography process, the doping of transistors in SRAM cells plays a crucial role in their behavior. Although the process aims for uniformity, variations in doping can occur due to the limitations of the fabrication equipment \cite{doping}. The machine moves along the x-axis, doping the transistors as it progresses from left to right or right to left. It then steps up and continues the doping process until all transistors are treated. These doping variations have an impact on the initial state and stability of the SRAM cells. The stability of an SRAM cell is typically evaluated using the concept of static noise margin (SNM), which represents the minimum noise voltage required to flip the state of the bitcell. In SRAM cell design, the width-to-length (W/L) ratios of the load transistors and access transistors are often set to be as close to 1.0 as possible. The ratio of the driver transistor's W/L to the access transistor's W/L is known as the cell ratio, which determines the cell's stability and size \cite{doping}. The doping variation directly affects the W/L ratio, potentially leading to variations in the SNM value. When considering the doping effect, transistors along the vertical axis experience more significant variations compared to adjacent transistors along the x-axis. These doping variations impact all SRAM-based PUFs to some extent. However, specific orientations, such as R270 and MY90, exhibit a distinctive negative bias pattern. In these orientations, the transistor arrangement becomes reversed or opposite to the baseline SRAM-based PUF's orientation. This is the main finding of our work.
\section{Conclusions} \label{sec:conclusion}
Our experiments revealed that the orientation affects the bias direction significantly. This observation could prove useful for designing smart error correction algorithms for SRAM-based PUFs. Regarding the evaluation characteristics of the SRAM-based PUFs, our findings are consistent with prior research, indicating that our SRAMs are representative. Nonetheless, our study presents a useful contribution to the field of PUF research and highlights the importance of careful physical implementation when designing SRAM-based PUFs.

\section{ACKNOWLEDGMENTS}
This work has been partially conducted in the project ``ICT programme'' which was supported by the European Union through the ESF. It was also supported by EU's Horizon 2020 R\&I programme under Grant Agreement No 872614 (SMART4ALL).

\bibliographystyle{IEEEtran}
\bibliography{output}

\begin{IEEEbiography}{Zain Ul Abideen}{\,}is currently pursuing his doctoral studies at Tallinn University of Technology in Estonia. His research work include hardware security and obfuscation of ASIC designs.
\end{IEEEbiography}

\begin{IEEEbiography}{Rui Wang}{\,}
is  working for Intrinsic ID as a technology engineer since 2017, focusing on various product research topics with PUF technology.
\end{IEEEbiography}

\begin{IEEEbiography}{Tiago Diadami Perez} {\,}received 
a Ph.D. degree from Tallinn University of Technology in Estonia. 
His current research interests include hardware security 
and IC implementation.
\end{IEEEbiography}

\begin{IEEEbiography}{Geert-Jan Schrijen} {\,}
joined the security group of Philips Research in Eindhoven where he worked on various topics, including PUFs, which resulted in the spin-off of Intrinsic ID in 2008. 
In 2011, Geert-Jan became responsible for all development and engineering work at Intrinsic ID in his role of VP Engineering. In 2016 he was appointed as CTO.
\end{IEEEbiography}

\begin{IEEEbiography}{Samuel Pagliarini} {\,} received the Ph.D. degree from Telecom ParisTech, Paris, France, in 2013. 
He is currently a Professor at Tallinn University of Technology in Estonia where he leads the Centre for Hardware Security.
\end{IEEEbiography}
\end{document}